\documentclass{aa}  
\usepackage{graphicx}
\usepackage{txfonts}

\begin{document}

\title{Nuclear $\gamma$-ray emission from very hot accretion flows}

\author{E. Kafexhiu\inst{1}
\and F. Aharonian\inst{1,2,3}
\and M. Barkov\inst{4,5,6}}

\institute{Max-Planck-Institut f\"ur Kernphysik, Saupfercheckweg 1, D-69117 Heidelberg, Germany
\and Dublin Institute for Advanced Studies, 31 Fitzwilliam Place, Dublin 2, Ireland
\and National Research Nuclear University MEPhI, Kashirskoje Shosse, 31, 115409 Moscow, Russia
\and Department of Physics and Astronomy, Purdue University, West Lafayette, IN 47907-2036, USA
\and Astrophysical Big Bang Laboratory, RIKEN, 351-0198 Saitama, Japan
\and Space Research Institute of the Russian Academy of Sciences (IKI), 84/32 Profsoyuznaya Str, Moscow, Russia, 117997}

\date{Received:  / Accepted: } 

\abstract{Optically thin accretion plasmas can reach ion temperatures $T_{\rm i} \geq 10^{10}$K and thus trigger nuclear reactions. Using a large nuclear interactions network, we studied the radial evolution of the chemical composition of the accretion flow toward the black hole and computed the emissivity in nuclear $\gamma$-ray lines. In the  advection dominated accretion flow (ADAF) regime, CNO and heavier nuclei are destroyed before reaching the last stable orbit. The overall luminosity in the de-excitation lines for a solar composition of plasma can be as high as few times $10^{-5}$ the accretion luminosity ($\dot{M}c^2$) and can be increased for heavier compositions up to $10^{-3}$. The efficiency of transformation of the kinetic energy of the outflow into high energy ($\geq 100$~MeV) $\gamma$-rays through the production and decay of $\pi^0$-mesons can be higher, up to $10^{-2}$ of the accretion luminosity. We show that in the ADAF model up to 15 percent of the mass of accretion matter can ``evaporate'' in the form of neutrons.}


\maketitle

\section{Introduction}
Hard X-ray emission from compact X-ray binary systems is a distinct signature of a hot plasma accreting onto a black hole. In an attempt to explain the X-ray emission from the stellar mass black hole Cygnus X-1,  \cite{Shapiro1976} developed an accretion disk model (the SLE model) whose inner region was considerably hotter and geometrically thicker than previous models. Accretion plasma is optically thin and is gas-pressure dominated. This model naturally yields a two-temperature plasma in which the electron temperature is about $k T_{\rm e}\sim 100$~keV and the ion temperature is one or two orders of magnitude larger. However,  later developments have shown that the SLE model is thermally unstable  \citep[see, e.g.,][]{Abramowicz1995, Abramowicz2000} and that the advection of heat plays an important role in the viscous and thermal stability of the plasma \cite[see, e.g.,][]{Ichimaru1977, paczinski1981, Abramowicz1988, narayan1993, Narayan1994a, Abramowicz1995, Chen1995}. 

Another group of very hot and optically thin accretion disk solutions is linked to the so-called advection-dominated accretion flows (ADAF) \cite{Narayan1994a}, and \cite{Narayan1995a,Narayan1995b}. These models are radiative inefficient because most of the plasma heat is advected into the black hole. In ADAF models, plasma is geometrically thick with a quasi-spherical flow. The electron temperature for ADAF can be as high as $k T_{\rm e}\sim 300$~keV, whereas ion temperature is close to the viral temperature and can be on the order of $k T_{\rm i} \sim 100$~MeV.

Nuclear reactions play an important role in such high ion temperature accretion plasmas. These reactions change the plasma chemical composition and produce a characteristic $\gamma$-ray emission, \citep{Aharonian1984, Aharonian1987, Yi&Narayan1997, kafexhiu2012}. Because of the high ion temperature, the nuclear spallation processes dominate. Therefore, plasma evolves toward a lighter composition. The $\gamma$-ray emission of the accretion plasma is governed by prompt nuclear de-excitation $\gamma$-ray lines. Their emissivities depend strongly on nuclear abundances. Therefore, their computation requires a detailed consideration of the chemical evolution of the accretion plasma \citep[see, e.g.,][]{kafexhiu2018b}.

When the ion temperature of plasma exceeds 10~MeV, it starts to produce $\pi^0$-mesons. Their decay produces an additional $\gamma$-ray emission component at higher energies \citep[see, e.g.,][]{dahlbacka1974, kolykhalov&sunyaev1979, bisnovatyi1980, giovannelli1982, dermer1986, madehavan1997, Yi&Narayan1997, oka2003, kafexhiu2018b}. Pions have a higher threshold energy than nuclear reactions. Pions are produced by energetic particles that populate the high energy tail of the plasma distribution function. Therefore, the $\gamma$-ray emission from $\pi^0$-meson decay is sensitive on the high energy tail of the plasma distribution function. Nuclear $\gamma$-ray lines, on the other hand, carry information about the plasma temperature because the most important part of their cross section is covered by the central part of the Maxwellian distribution function. Thus, the combined information about the prompt $\gamma$-ray lines and the $\pi^0 \to 2 \gamma$  emissions should allow reconstruction of the plasma energy distribution function for a wide energy range.
 
In this paper, we present the results of our study of the $\gamma$-ray emission related to two-temperature accretion plasma within the framework of the ADAF and SLE models. For this task, we utilize a massive nuclear reaction network and the recent parametrizations of $\pi$-meson production cross sections at low (close to the kinematic threshold) energies. We study the radial chemical evolution of the accretion flow and compute the resulting $\gamma$-ray emission. We also discuss the efficiency of evaporation of free neutrons from the disk and the maximum amount of the secondary Li, Be, and B nuclei produced in the ADAF accretion plasma.

\section{Radial evolution of the chemical composition}   

\begin{figure*}
\centering
\includegraphics[scale=0.45]{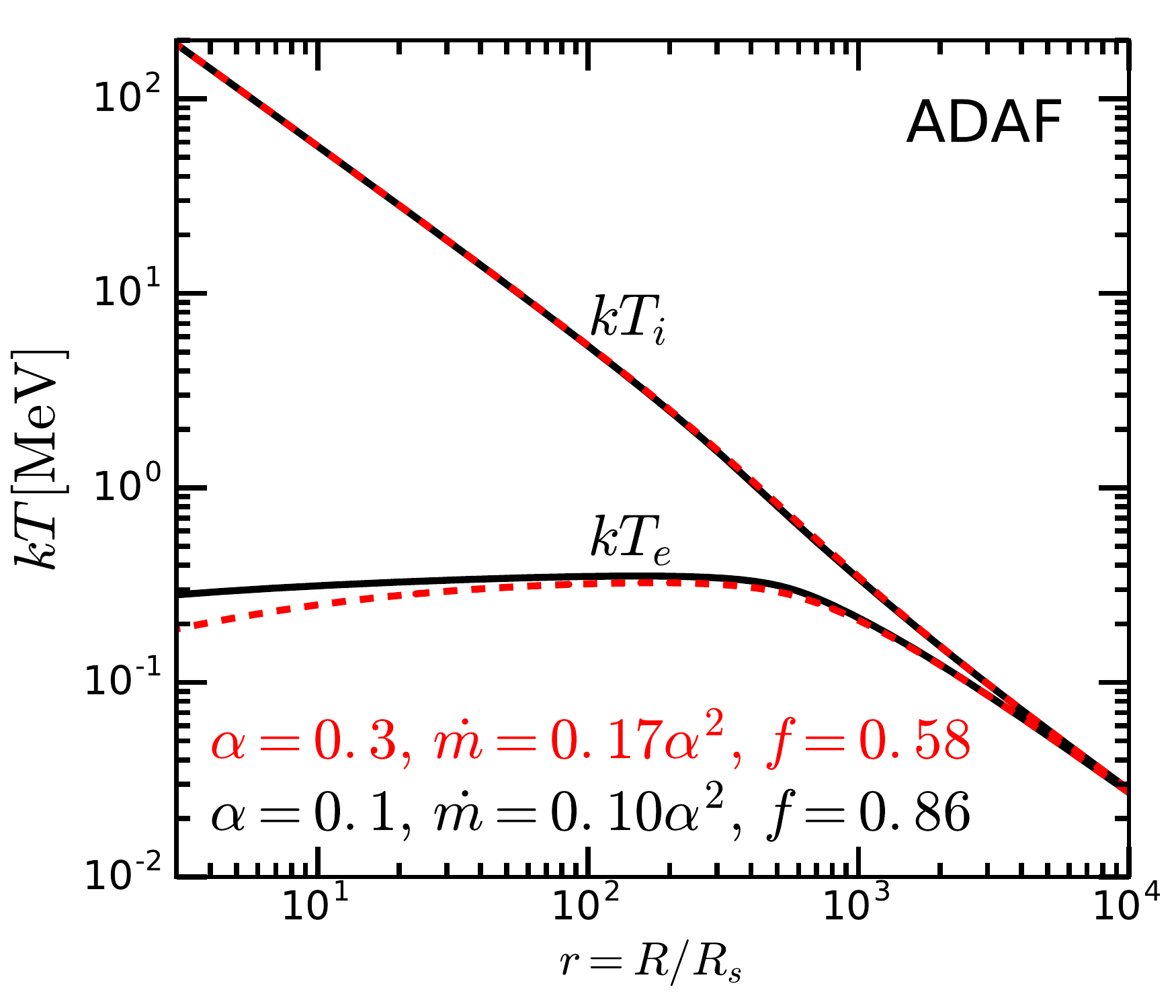}
\includegraphics[scale=0.45]{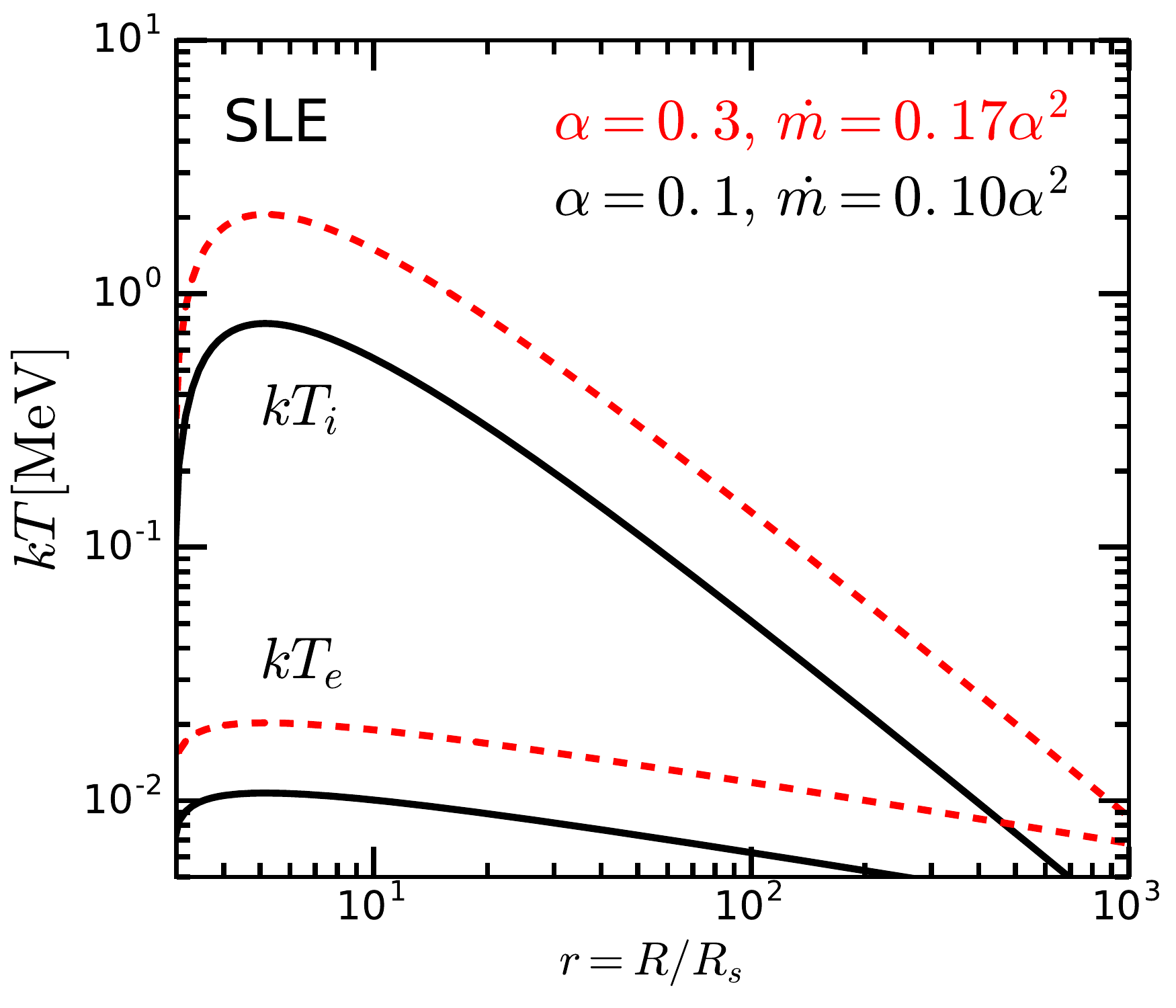}\\
\includegraphics[scale=0.45]{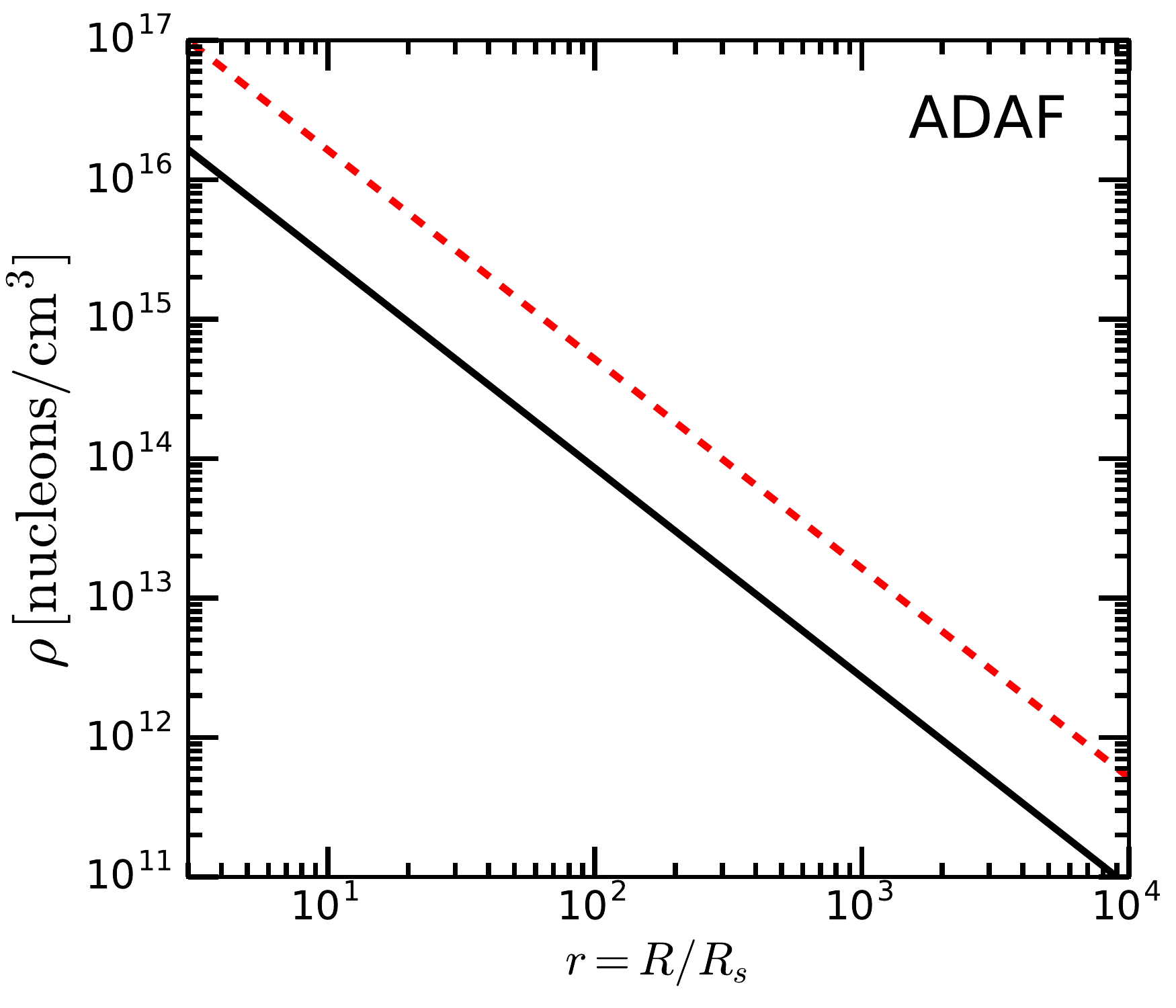}
\includegraphics[scale=0.45]{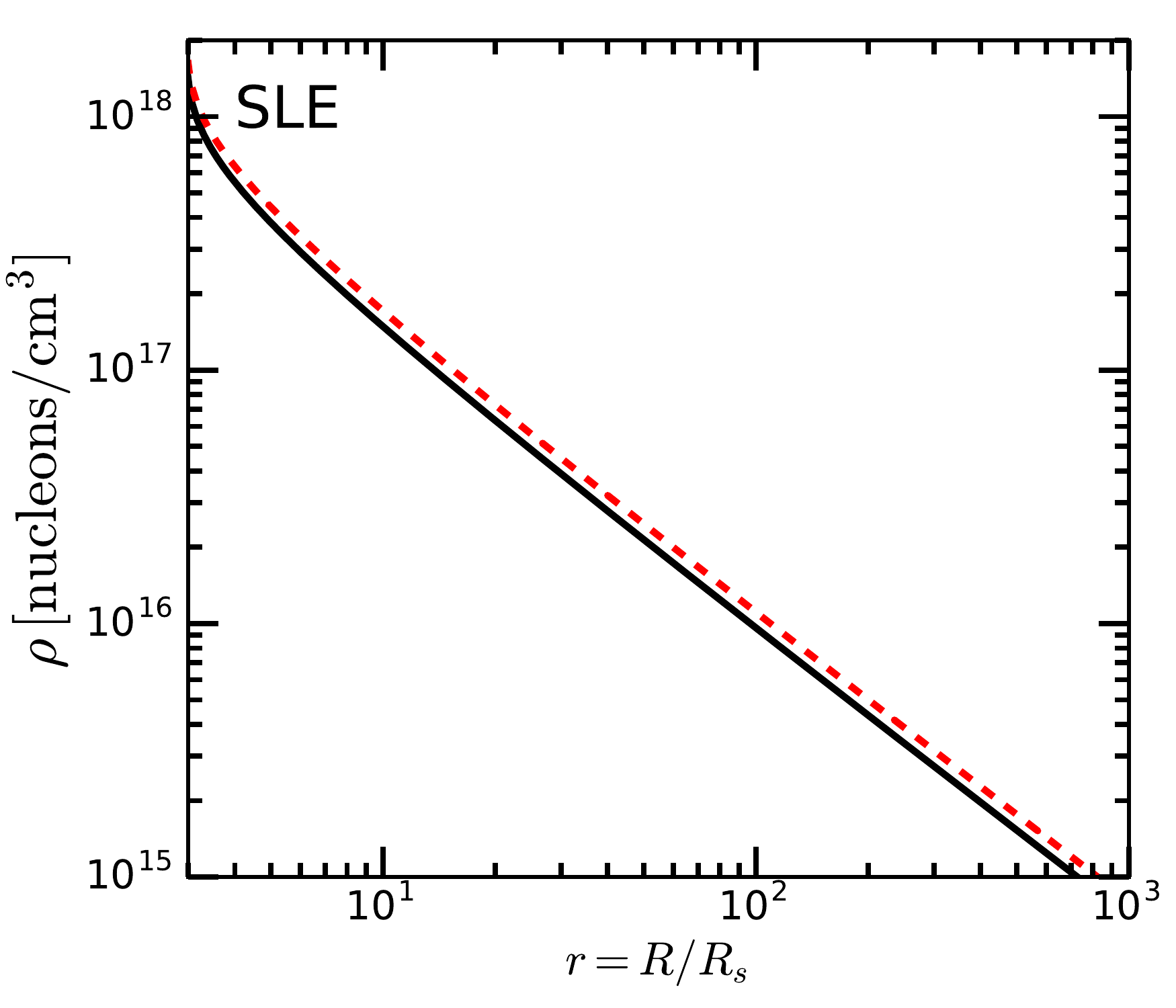}
\caption{Radial profiles of the temperature (top) and nucleon number density (bottom) profiles for the ADAF (left) and SLE (right) accretion disk models. The accretion disk parameters are $\alpha=0.1$ and $\dot{m}=0.1\alpha^2$ (black lines) and $\alpha=0.3$ and $\dot{m}=0.17\alpha^2$ (dashed red lines); the mass $m=10$ and $\beta=0.5$.} 
\label{fig:TempDens}
\end{figure*}

\subsection{Nuclear reaction network}

Low density accretion plasma, unlike, for example the central core of a star, is dominated by binary nuclear collisions. For a binary reaction $i+j\to l +...$, the production rate per unit volume of the nucleus $l$ is written as \citep[see, e.g.,][]{Fowler1967}\begin{equation} \label{eq:prodRate}
\dot{n}_l = \frac{n_i\,n_j}{1+\delta_{ij}}\,\left\langle \sigma v\right\rangle_{ij}^l.
\end{equation}
In this equation $n_i$, $n_j$ and $n_l$ are the number densities of species $i$, $j,$ and $l$, respectively; $\delta_{ij}=1$ if nucleus $i$ and $j$ are identical, and zero otherwise. The $\left\langle \sigma v\right\rangle_{ij}^l$ is the thermal reaction rate, which is computed from
\begin{equation}
\left\langle \sigma v\right\rangle_{ij}^l(T) = \sqrt{\frac{8}{\pi\,\mu_{ij}\,(k T)^3}} \int\limits_0^\infty \sigma_{ij}^l(E)\,E\,\exp\left(-\frac{E}{k T}\right)\, dE,
\end{equation}
where, $\mu_{ij}=m_i\,m_j/(m_i+m_j)$ is the reduced mass of the two incoming nuclei, $\sigma_{ij}^l$ is the $i+j\to l+...$ reaction cross section,  $k T$ is the temperature in units of energy, and $E$ is the collision energy.

Equation~(\ref{eq:prodRate}) is more convenient to solve in terms of the mass fraction abundance. For an element $i$, the mass fraction abundance is defined as $X_i =A_i\,n_i/\rho $, where $A_i$ is the mass number, $n_i$ is its number density, and $\rho=\sum A_i\, n_i$ is the nucleon number density of the plasma. If the accretion plasma is initially composed of $N$ nuclei, then we have to sum up over all possible channels that produce a given nucleus $l$. Moreover, nuclear reactions can produce nuclei that were not initially present in the plasma. Therefore, we have to consider all possible nuclear species that can be produced during the plasma lifetime and all their most important nuclear reaction channels. This system forms a so-called nuclear reaction network. For calculating the cross sections $\sigma_{ij}^l$ we use the nuclear code TALYS \citep[][]{talys} and experimental data, for more details see \cite{kafexhiu2012}. We note that the recent version of the code TALYS can calculate cross sections for energies up to 1~GeV. This has allowed us to extend the computation of the reaction rates $\left\langle \sigma\, v\right\rangle_{ij}^l$ for temperatures on the order of $10^{12}$~K. In addition, the nuclear reaction network considers $p+p\to p+n+\pi^+$ and $p+p\to D+\pi^+$ channels, which can significantly change the abundances of neutrons and deuterium in case of a proton plasma with temperatures $k T_{\rm i}>10$~MeV; see \cite{kafexhiu2018b} for more details.

The solutions of the nuclear network are obtained by solving the following system of differential equations:
\begin{equation}
\frac{dX^l}{dR} = \sum\limits_{i,j}^N \frac{\rho}{(1+\delta_{ij})\,v}\,\left(\frac{A_l}{A_i\,A_j}\right)\,\left\langle \sigma v\right\rangle_{ij}^l\, X^i\,X^j.
\end{equation}
We converted the time derivative into the radial derivative using $\dot{X}^l=v\,dX^l/dR$, thus solving $X^l$ as a function of the disk radius. 

In addition, our nuclear reaction network considers reactions $i+j\to l^*+...,$ which produce nucleus $l$ in different excited states that lead to different prompt $\gamma$-ray lines emission. The cross sections for the most important prompt nuclear $\gamma$-ray lines are calculated using TALYS code and experimental data \citep[see, e.g.,][]{Kozlovsky2002, Murphy2009}. The number of $\gamma$-rays produced per unit volume $\dot{n}_\gamma$ for a specific $l^*$ transition, is calculated using eq.~(\ref{eq:prodRate}). The reaction rate $\left\langle \sigma v\right\rangle_{ij}^{l^*}$ is calculated from the specific $\sigma_{ij}^{l^*}$ cross section. Since the accretion plasma temperatures are $kT>1$~MeV, the  Doppler broadening from thermal motion of $l^*$ is larger than the natural width of the prompt $\gamma$-ray line. Therefore, the $\gamma$-ray line profile is well described by a Gaussian and its emissivity is given by
\begin{equation}
\phi_\gamma(E_\gamma,r) = \frac{\dot{n}_\gamma(r)}{\sqrt{2\pi\,\sigma_G^2(r)}}\,\exp\left[-\frac{(E_\gamma-E_\gamma^0)^2}{2\,\sigma_G^2(r)}\right].
\end{equation}
In this equation, $E_\gamma^0$ is the central $\gamma$-ray energy obtained by the energy difference of the two transition levels of the nucleus, $\sigma_G=E_\gamma^0\,\sqrt{k T/m_lc^2}$ is the Gaussian broadening, and $m_l$ is the mass of the excited nucleus $l^*$.

With the exception of  nuclear lines, nuclear collisions produce a $\gamma$-ray continuum. For $E_\gamma\gg10$~MeV, the nuclear continuum is composed of mainly nuclear bremsstrahlung and the production and decay of the $\pi^0$-mesons. Calculation of the emissivity $\phi_\gamma$ for these channels is described in \cite{kafexhiu2018b}. To compute the differential luminosity of the disk, we have to integrate $\phi_\gamma(E_\gamma,R)$ over the entire volume of the disk, i.e., $\Phi_\gamma(E_\gamma) =\int\, \phi_\gamma(E_\gamma,R)\; 2H\, 2\pi R\, dR$.

\subsection{Principal parameters of the accretion disk \label{sec:Parameters} }

Accretion disk is described by the following set of key parameters: (1) the viscosity parameter $\alpha$; (2) the scaled black hole mass $m=M/M_\odot$, where $M_\odot$ is the solar mass; (3) the scaled accretion rate $\dot{m}=\dot{M}/\dot{M}_{edd}$, where $\dot{M}_{edd}$ is the so-called Eddington accretion rate; (4) the ratio of the gas pressure to the total pressure $\beta$; (5) the adiabatic index $\gamma$; (6) the viscous heat parameter $\delta$, which gives the ratio of electron heating to total heating; and (7) the advection parameter $f$, which gives the ratio of the advection heating to the total heating \citep[see, e.g.,][]{Narayan1995b}. Apart from the system specific parameters such as the $m$ and $\dot{m}$, most of the other parameters are constrained from numerical simulations or are limited by other parameters in a self-consistent way.

By fixing the values of $m$, $\dot{m}$, $\alpha,$ and $\beta$, we can calculate the physical quantities of the plasma such as  ion and electron temperatures ($T_i$ and $T_e$), plasma nucleon number density ($\rho$), plasma radial speed ($\varv$), and the disk height ($H$), which are required to solve the nuclear reaction network. For the ADAF accretion, the above quantities are calculated from \citep[see, e.g.,][]{Narayan1995b}

\begin{equation}
\begin{array}{l}
T_i+1.08\,T_e=6.66\times 10^{12}\,\beta\,c_3\,r^{-1}\;{\rm K},\\
\rho = 2.28\times 10^{19}\,\alpha^{-1}\,c_1^{-1}\,c_3^{-1/2}\,m^{-1}\, \dot{m}\, r^{-3/2}\;{\rm nucleons/cm^{3}},\\
\varv = -2.12\times 10^{10}\,\alpha\,c_1\,r^{-1/2}\;{\rm cm\,s^{-1}},\\
H = (2.5\,c_3)^{1/2}\, R.
\end{array}
\end{equation}
In this case, $r=R/R_s$, where $R$ is the disk radius and $R_s$ is the Schwarzschild radius. $c_1=g(\alpha,\epsilon')\times(5+2\epsilon')/3\alpha^2$ and $c_3=2c_1/3$, where, $\epsilon'=\epsilon/f=f^{-1}\,(5/3-\gamma)/(\gamma-1)$ and $g(\alpha,\epsilon')=\left[1+18\alpha^2/(5+2\epsilon')^2\right]^{1/2} -1$. The plasma ion and electron temperature profiles are computed from the local balance equations $q^+=q^{\rm adv} + q^{\rm ie}$ for ions and $q^{\rm ie} = q^-$ for electrons. The quantity $q^+$ is the plasma heating energy rate due to the viscous dissipation, $q^{\rm adv}=f\, q^+$ is the cooling rate due to advection, and $q^{\rm ie}$ is the cooling rate due to the Coulomb coupling with electrons. Electrons, on the other hand, are heated through  Coulomb collisions with ions $q^{\rm ie}$ and are cooled by synchrotron, bremsstrahlung, and Comptonization, the sum of which gives $q^-$; for full details see \cite{Narayan1995b}. The solution of the local balance equations determines the value of $f$.

For the SLE accretion disk model the above quantities are computed from \citep[see][]{Shapiro1976}
\begin{equation}
\begin{array}{l}
T_i = 5\times 10^{11}\, M_*^{-5/6}\,\dot{M}_*^{5/6}\,\alpha^{-7/6}\,\mathcal{F}^{5/6}\, r_*^{-5/4} \;{\rm K},\\
T_e = 7\times 10^{8}\,\left(M_*\,\dot{M}_*^{-1}\,\alpha^{-1}\,\mathcal{F}^{-1}\right)^{1/6}\, r_*^{1/4} \;{\rm K},\\
\rho = 3\times 10^{19}\, M_*^{-3/4}\,\dot{M}_*^{-1/4}\,\alpha^{3/4}\,\mathcal{F}^{-1/4}\, r_*^{-9/8}\;{\rm nucleons/cm^{3}},\\
H = 10^5\, M_*^{7/12}\,\dot{M}_*^{5/12}\,\alpha^{-7/12}\,\mathcal{F}^{5/12}\, r_*^{7/8}\;{\rm cm}.\\
\end{array}
\end{equation}
In this equation, $M_*=M/3$, $\dot{M}_*=\dot{M}/\left(10^{17}\,{\rm g\,s^{-1}}\right)$, $r_*= 2\,r$ and $\mathcal{F}=1-\sqrt{3/r}$. The radial velocity is obtained from the continuity equation and is given by $\varv=-\dot{M}/(2 H\, 2\pi R \,\rho\, m_u)$, where $m_u$ is the nucleon mass. We note that $H$ is half of the disk height.

\subsection{Radial chemical evolution of the accreting flow\label{subsec:chemical}}

In this section,  we apply the nuclear reaction network to the ADAF and SLE disk models. Throughout this article, we consider a 10 solar mass black hole,  $m=10$, that is typical for binary systems and assume an equipartition magnetic field with $\beta=0.5$. The ADAF accretion disk exists below some critical values of the accretion rate and plasma density. Therefore, we choose $\alpha=0.1$ and $\dot{m}=0.1\times\alpha^2=10^{-3}$ far from the critical values and  $\alpha=0.3$ and $\dot{m}=0.17\times\alpha^2 = 1.53\times10^{-2}$ near the critical value. The ADAF and SLE temperature and density profiles for this set of parameters are shown in Fig.~\ref{fig:TempDens}. In addition, we consider two initial chemical compositions: a solar composition realized when the donor is a normal star and a heavy composition expected when the donor is a Wolf-Rayet type star. More specifically we consider H, ${\rm ^4He}$, ${\rm ^{12}C,}$ and ${\rm ^{16}O}$ with mass fractions $X_p=X_\alpha=0.3$ and $X_{C12}=X_{O16}=0.2$.

\begin{figure*}
\centering
\includegraphics[scale=0.45]{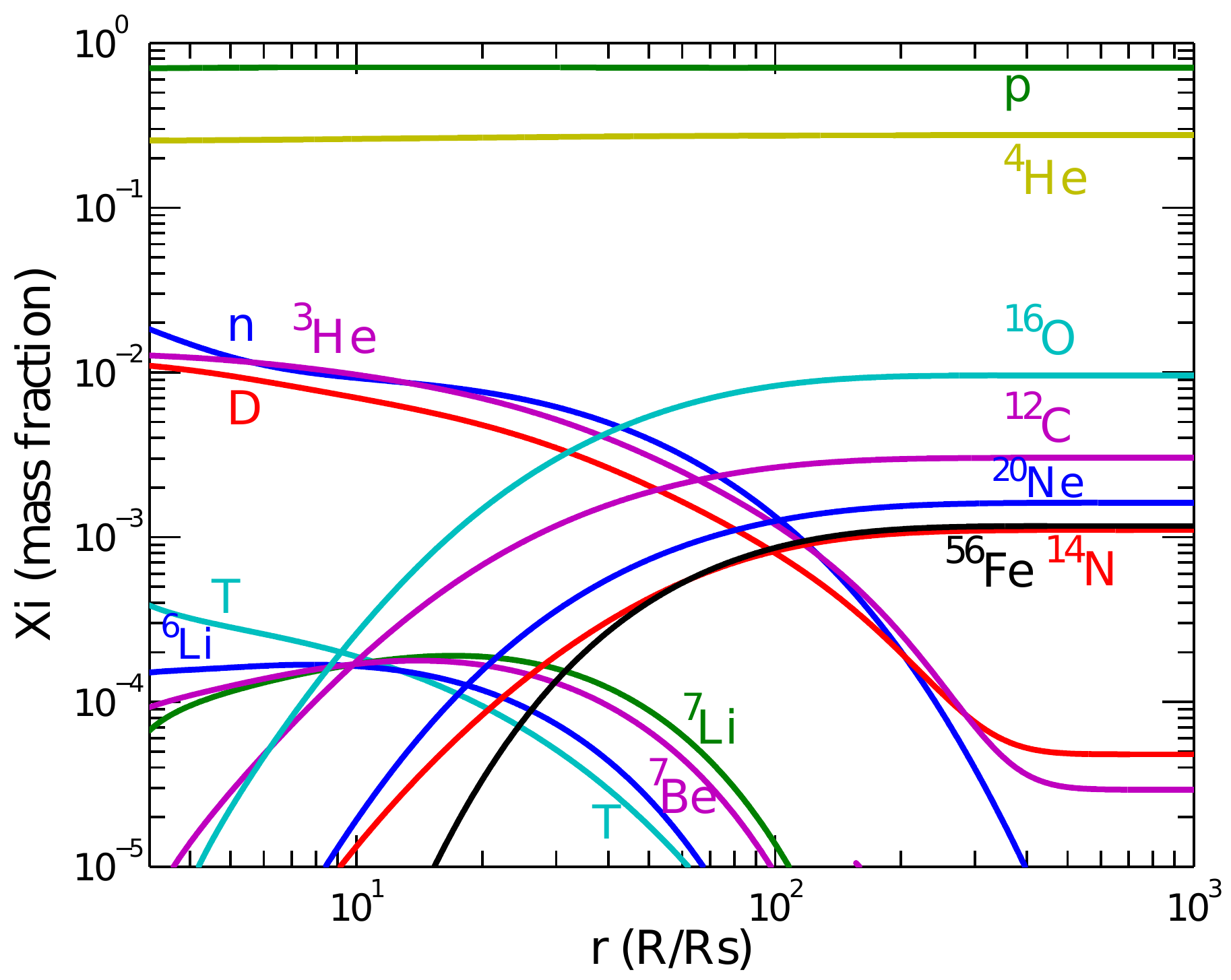}
\includegraphics[scale=0.52]{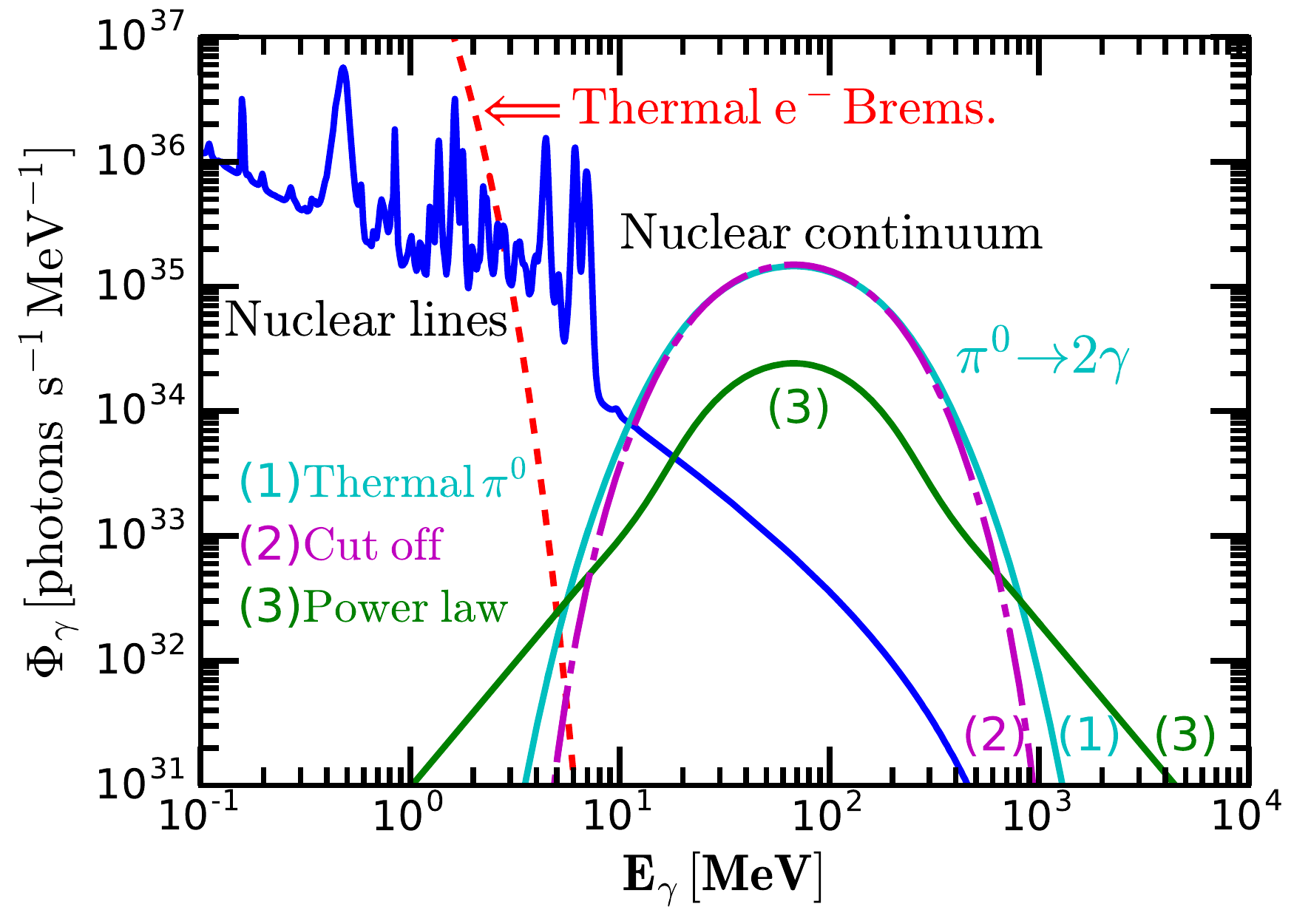}
\caption{Radial chemical evolution (left) and the disk $\gamma$-ray luminosity (right). The results are calculated for the ADAF model for parameters $m=10$, $\alpha=0.1$, $\dot{m}=10^{-3}$, $\beta=0.5,$ and the initial solar composition. The left panel shows the evolution of elements. The light elements and the initial nuclei with mass fraction $X_i>10^{-3}$ are explicitly shown. The right panel shows the disk differential luminosities (in units of $\rm photons/cm^2 s MeV$) of nuclear $\gamma$-rays consisting of resolved $\gamma$-ray lines and the continuum composed of superposition of broad nuclear lines, neutron capture radiation, and nuclear bremsstrahlung emission). The curves denoted as (1), (2), and (3) show the disk differential $\pi^0$-decay $\gamma$-ray luminosities: curve (1) corresponds to the nominal Maxwellian distribution of particles, curve (2) corresponds to a Maxwellian distribution with a sharp cutoff at $E_k=4 k T_{\rm i}$, whereas curve (3) corresponds to the Maxwellian distribution that has a power-law tail with an index $p=2.0$ for $E_k>4\,k T_{\rm i}$. The luminosity of the thermal electron bremsstrahlung (red dashed line) is also shown. The advection factor for this model is $f=0.86$ and the accretion luminosity is $L_{acc}=1.4\times10^{36}~\rm{erg~s^{-1}}$. The integrated luminosity that the electrons radiate away through bremsstrahlung is $L_e=1.9\times 10^{35}~\rm{erg~s^{-1}}$, the luminosity in the the nuclear lines and continuum is  $L_N =3.3\times 10^{31}~\rm{erg~s^{-1}}$, and the luminosity of radiation from decays of the secondary $\pi^0$-mesons is $L_\pi =9.3\times 10^{33}~\rm{erg~s^{-1}}$.} 
\label{fig:ADAF-specSOLAR1}
\end{figure*}

\begin{figure*}
\centering
\includegraphics[scale=0.45]{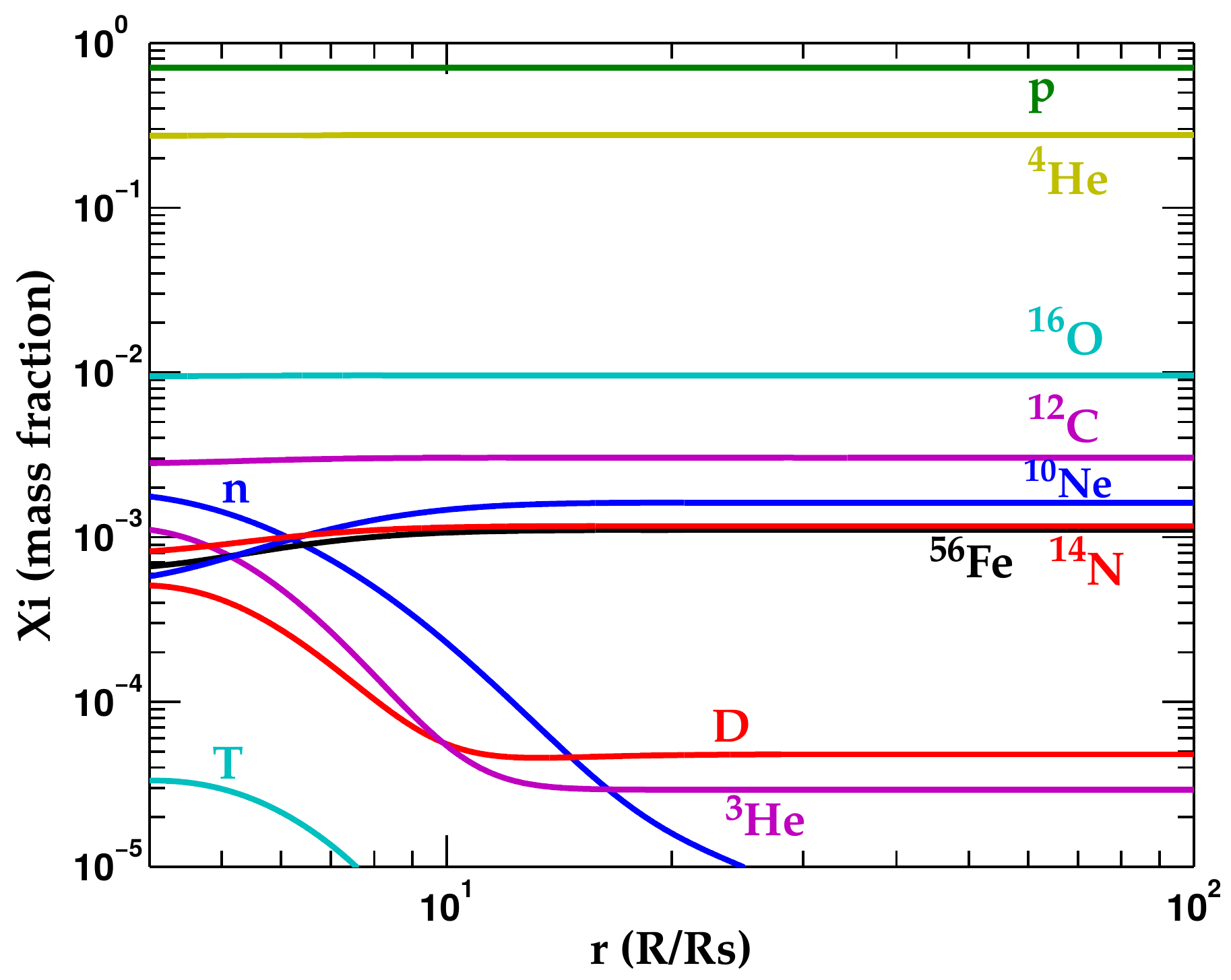}
\includegraphics[scale=0.52]{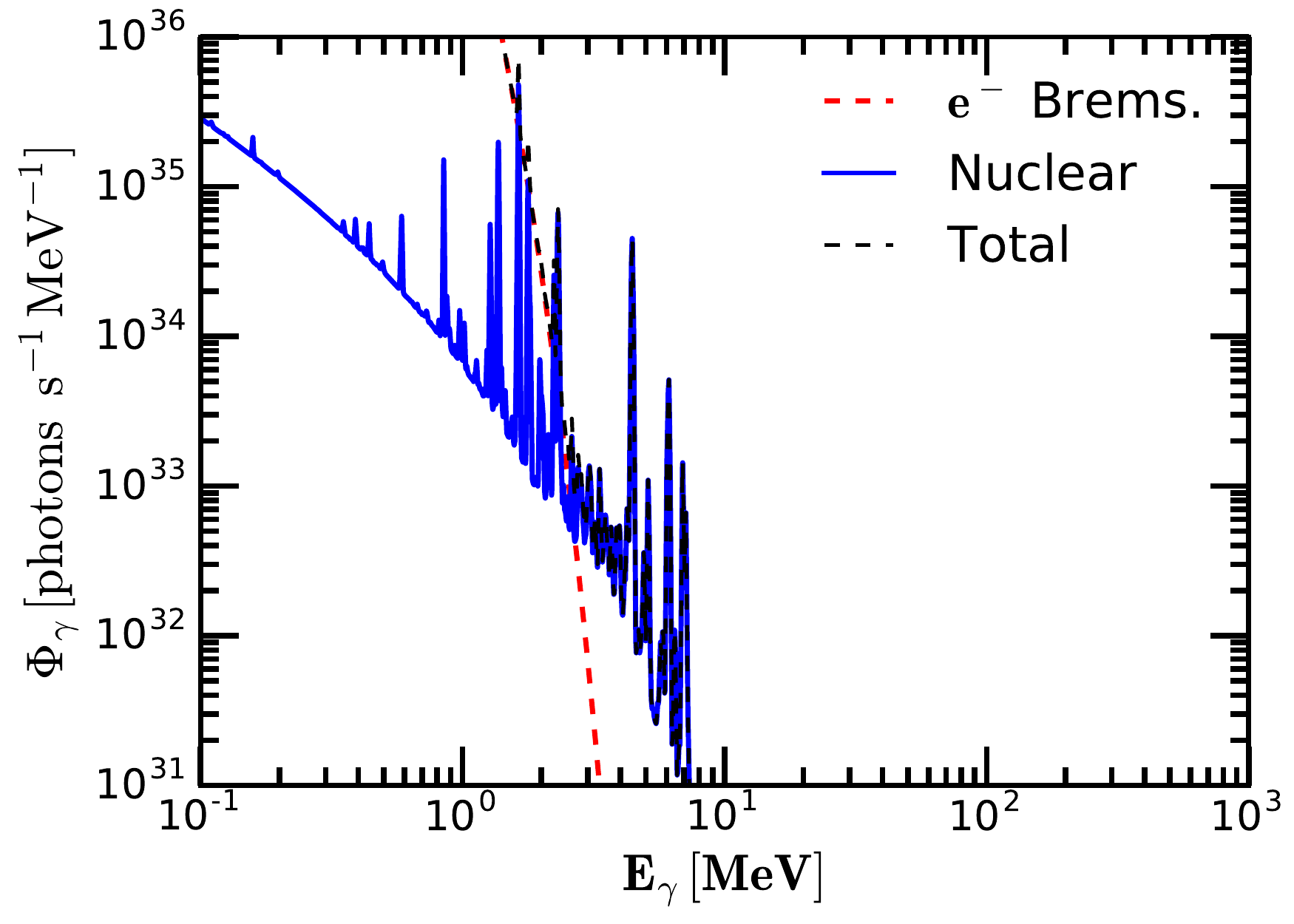}
\caption{Same as Fig.~\ref{fig:ADAF-specSOLAR1} but for the SLE disk model for the parameters $m=10$, $\alpha=0.1$, $\dot{m}=10^{-3}$, $\beta=0.5$. The accretion luminosity for this model is $L_{acc}=1.4\times 10^{36}~\rm{erg~s^{-1}}$. The integrated luminosity that electrons radiate away through bremsstrahlung is $L_e =9.4\times 10^{35}~\rm{erg~s^{-1}}$, the luminosity due to nuclear interactions is $L_N =5.6\times 10^{30}~\rm{erg~s^{-1}}$, and the luminosity of the $\pi^0$-decay $\gamma$-rays  is negligible.} 
\label{fig:SLE-spec}
\end{figure*}

\begin{figure*}
\centering
\includegraphics[scale=0.45]{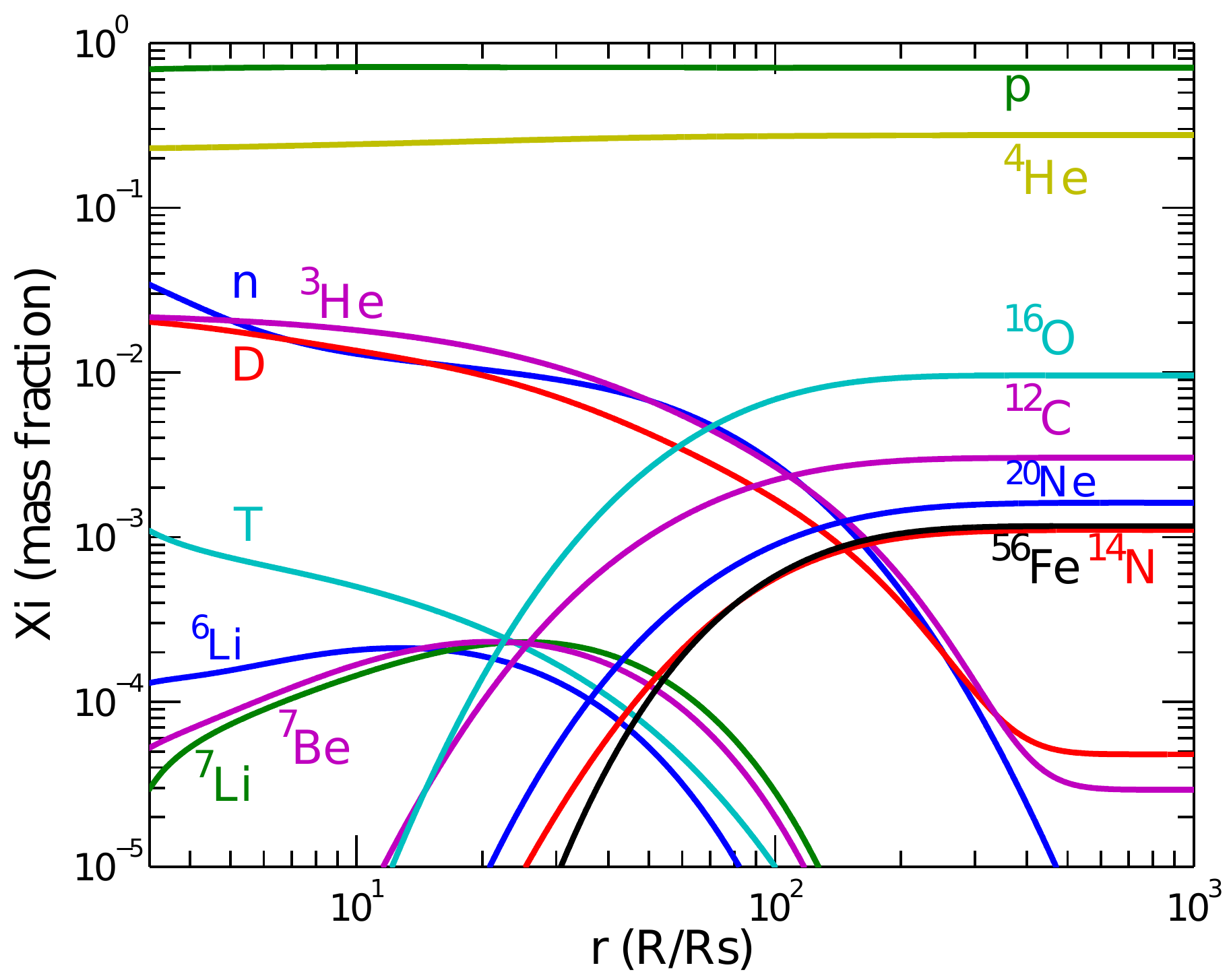}
\includegraphics[scale=0.52]{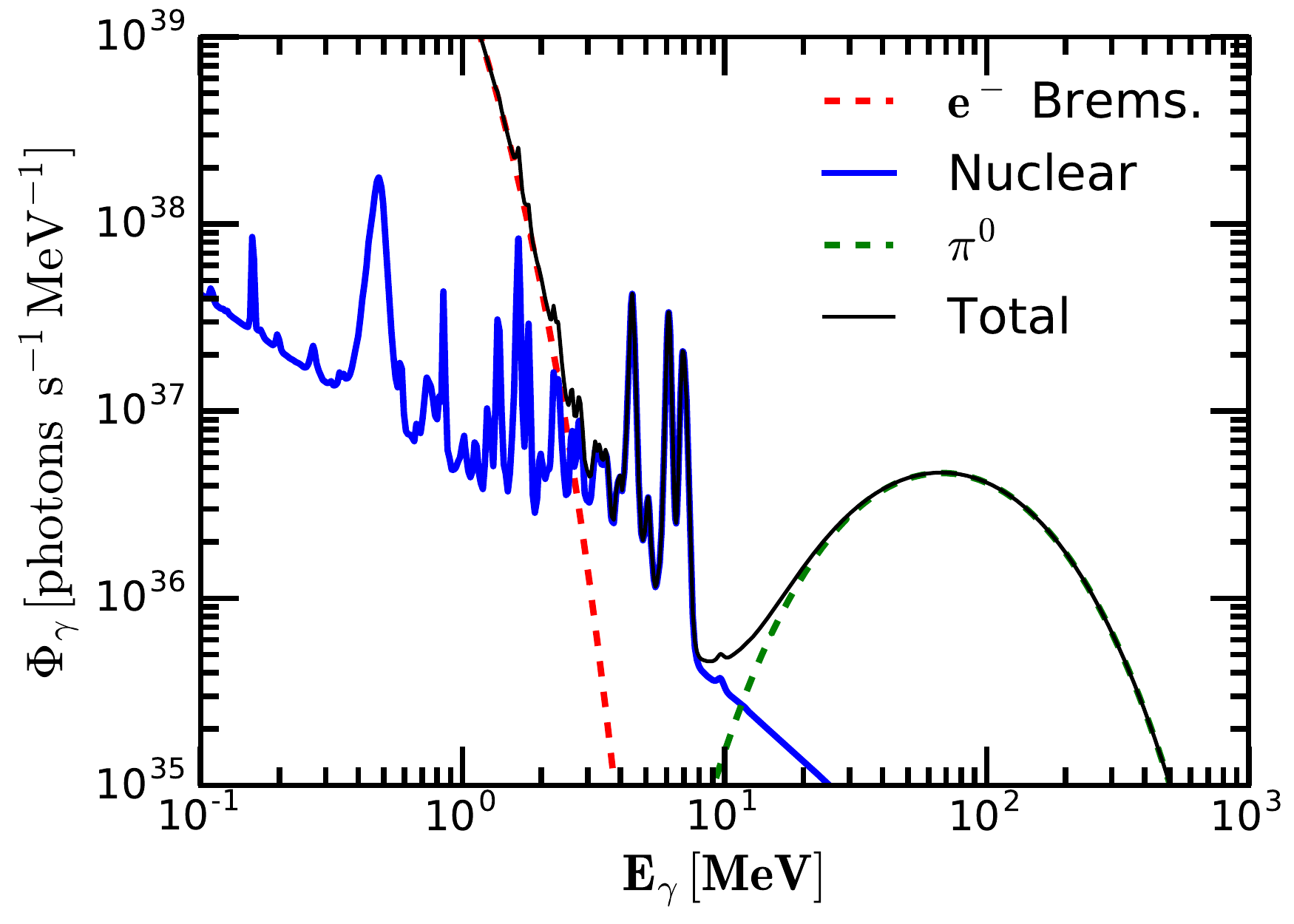}
\caption{Same as Fig.~\ref{fig:ADAF-specSOLAR1}, but for the parameters  $m=10$, $\alpha=0.3$, $\dot{m}=1.7\times\alpha^2=1.53\times10^{-2}$, $\beta=0.5$. The left panel shows the evolution of some important nuclei. The right panel shows the differential photon luminosity of nuclear $\gamma$-rays (blue curve) and the $\pi^0$-decay $\gamma$-rays (cyan line).  The luminosity of the thermal electron bremsstrahlung is also shown (red dashed curve). The advection factor for this model is $f=0.58$ and the accretion luminosity is $L_{acc}=2.1\times10^{37}~\rm{erg~s^{-1}}$. The total luminosity in the nuclear $\gamma$-ray lines and continuum is $L_N =1\times 10^{33}~\rm{erg~s^{-1}}$ and the luminosity radiated through the thermal $\pi^0$-meson production is $L_\pi =3.3\times 10^{35}~\rm{erg~s^{-1}}$.} \label{fig:ADAF-specSOLAR2}
\end{figure*}

\begin{figure*}
\centering
\includegraphics[scale=0.45]{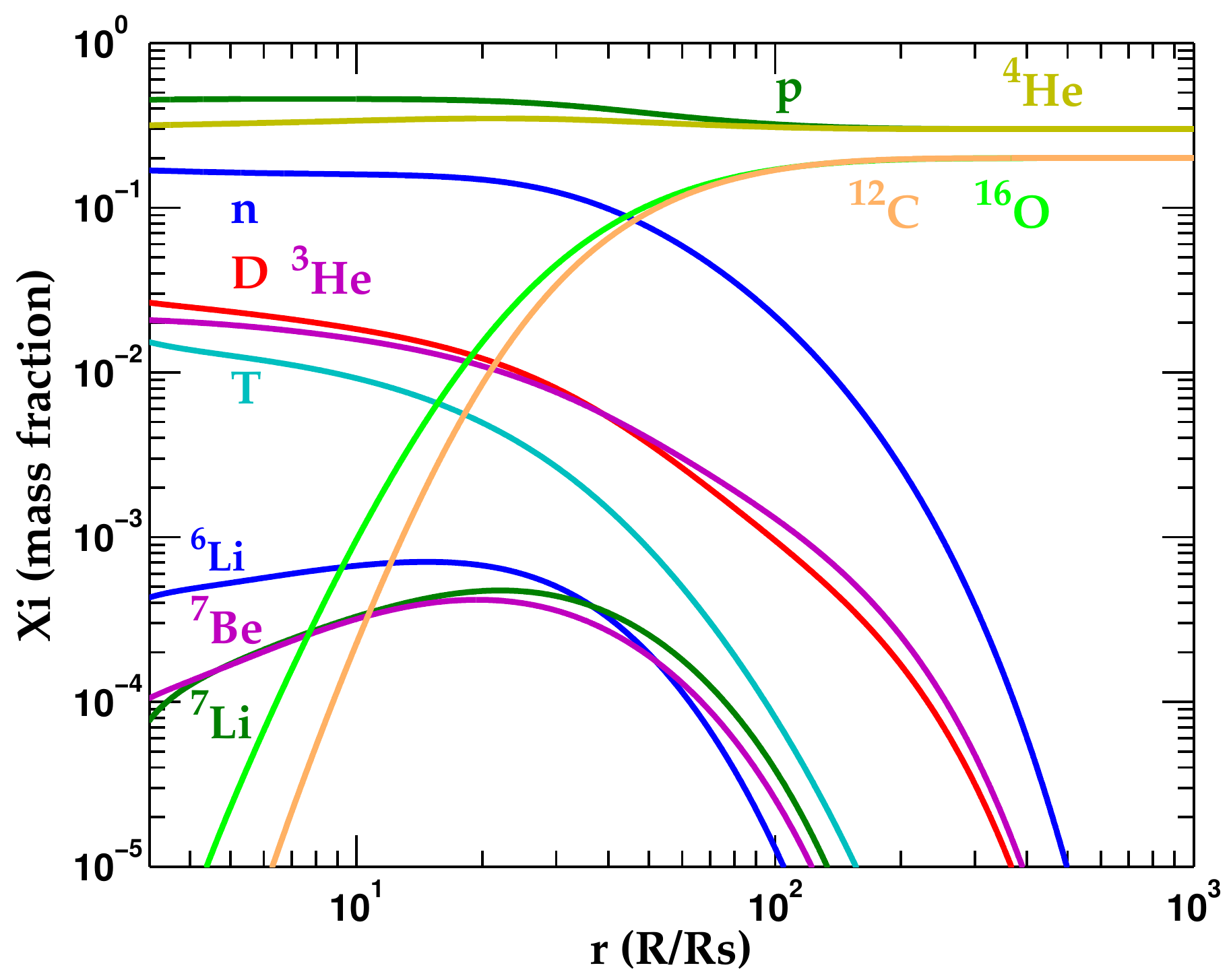}
\includegraphics[scale=0.52]{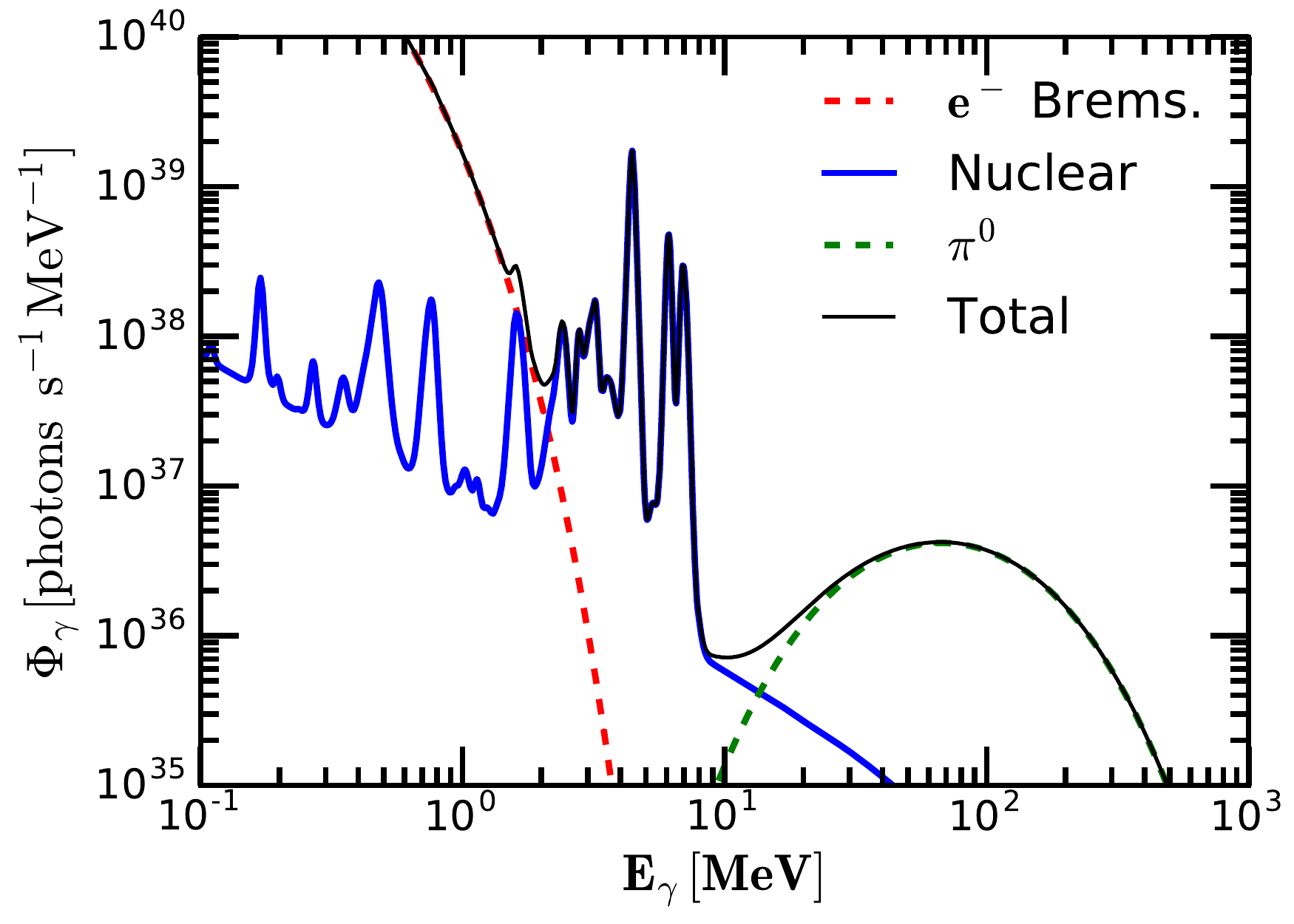}
\caption{Same as Fig.~\ref{fig:ADAF-specSOLAR2} but for the initial composition with $X_p=X_\alpha=0.3$ and $X_{C12}=X_{O16}=0.2$. The total luminosity in nuclear lines and continuum is $L_N =8.8\times 10^{33}~\rm{erg~s^{-1}}$, while the luminosity in $\pi^0$-decay $\gamma$-rays is $L_\pi =2.9\times 10^{35}~\rm{erg~s^{-1}}$.}\label{fig:ADAF-specCO}
\end{figure*}

The radial chemical evolution of the ADAF and SLE accretion disks are shown in Figs~\ref{fig:ADAF-specSOLAR1} -- \ref{fig:ADAF-specCO}. Figure~\ref{fig:ADAF-specSOLAR1} and \ref{fig:SLE-spec} show the results for a solar initial composition and $\alpha=0.1$ and $\dot{m}=10^{-3}$ parameters. Figure~\ref{fig:ADAF-specSOLAR2} and \ref{fig:ADAF-specCO} show the same results using ADAF accretion parameters near the critical values\footnote{The ADAF solution is valid for accretion rates $\le 0.17\times\alpha^2$ \citep[see, e.g.,][]{Abramowicz1995,Abramowicz2000}. For higher accretion rates the accretion regime switches to the relatively cold and thin solution of the \cite{ShakuraSunyaev1973} disk.} with $\alpha=0.3$ and $\dot{m}=1.53\times10^{-2}$. The results in Figure~\ref{fig:ADAF-specSOLAR2} are calculated for a solar initial composition, whereas Fig.~\ref{fig:ADAF-specCO} corresponds to the heavier (H-He-C-O) initial composition.

\section{$\gamma$-ray emission of the accretion disk}

\subsection{$\gamma$-ray luminosities \label{sec:Luminos}}

The differential $\gamma$-ray production rates ("photon luminosity" in units of $\rm photons/(MeV \, s)$) calculated for the ADAF and SLE accretion disk models are shown in Figs~\ref{fig:ADAF-specSOLAR1} -- \ref{fig:ADAF-specCO}. The key model parameters used in calculations are described in Sec.~\ref{subsec:chemical}.

The $\gamma$-ray luminosity of the accretion disk scales as $L\propto \rho^2\times R^3\propto m\,\dot{m}^2\,\alpha^{-2}$, where $\rho$ is the plasma density. We note, however, that this scaling is valid for accretion parameters far from their critical values. We let $L_{\rm acc}=\dot{\rm M}\,c^2= \dot{m} \times L_{\rm Edd}$ be the accretion luminosity, where $L_{\rm Edd}$ is the so-called Eddington luminosity. Let us denote with $L_{\rm N}$ the $\gamma$-ray luminosity linked to the nuclear reactions (including the prompt de-excitation lines and continuum) and with $L_\pi$ the $\gamma$-ray luminosity from the $\pi^0$-meson decay. For an ADAF accretion plasma with a solar initial composition the calculations give (i) far from the critical regime (i.e., $\dot{m} \ll 0.17 \alpha^2$ and $\alpha=0.3$),

\[
\begin{array}{cl}
L_N & \approx  3.3\times 10^{31} \left(\frac{m}{10}\right) \left(\frac{\dot{m}}{10^{-3}}\right)^2 \left(\frac{\alpha}{0.1}\right)^{-2}\,\rm{erg~s^{-1}},\\
L_\pi & \approx 9.3\times 10^{33} \left(\frac{m}{10}\right) \left(\frac{\dot{m}}{10^{-3}}\right)^2 \left(\frac{\alpha}{0.1}\right)^{-2}\,\rm{erg~s^{-1}},\\
\end{array}
\]

\noindent (ii) near the critical regime, where the luminosities approach their maximum but do not exceed

\[
\begin{array}{l}
L_N = 1\times 10^{33} \left(\frac{m}{10}\right)\,\rm{erg~s^{-1}},\\
L_\pi = 3\times 10^{35} \left(\frac{m}{10}\right)\,\rm{erg~s^{-1}}.\\
\end{array}
\]

For an initial heavy composition plasma with $X_p=X_\alpha=0.3$ and $X_{C12}=X_{O16}=0.2$, these upper limits are changed to
\[
\begin{array}{l}
L_N = 9\times 10^{33} \left(\frac{m}{10}\right)~\rm{erg~s^{-1}},\\
L_\pi = 3\times 10^{35} \left(\frac{m}{10}\right)~\rm{erg~s^{-1}}.\\
\end{array}
\]

For the SLE model, the luminosity radiated by nuclei for an initial solar composition and $\alpha=0.3$ and $\dot{m}= 0.17 \alpha^2$, is written as
\[
L_N = 6\times 10^{32} \left(\frac{m}{10}\right)~\rm{erg~s^{-1}},
\]

while in the initially heavy composition plasma,
\[
L_N = 1\times 10^{34} \left(\frac{m}{10}\right)~\rm{erg~s^{-1}}.
\]

In the SLE model, the luminosity of $\pi^0$-decay $\gamma$-rays is negligible because the relatively low ion temperature, $k T_{\rm i} \leq 10$~MeV.

\subsection{Modified high energy tail of the Maxwellian distribution function}

Generally, for the thermal plasma, the energy distribution of particles is silently assumed to be described by the Maxwellian distribution. However, this could be not a correct representation as long as it concerns the sub-relativistic ion component of the accretion plasma. The particle distribution well above the temperature can be suppressed compared to the Maxwellian distribution if the relaxation time of the tail $E \gg kT$ supported by the binary elastic scatterings would be longer than the accretion time. In this case, the high energy tail would be underdeveloped and correspondingly, the $\gamma$-ray luminosity suppressed, especially at energies well beyond 100~MeV \citep[]{AharonianAtoyan1983}. On the other hand, even a slight acceleration of ions, for example due to plasma instabilities, can produce a suprathermal component well above the "nominal" tail of the Maxwellian distribution.  Thus, the shape of the high energy tail of the accretion plasma distribution function can have dramatic effects on the pion production rate \citep{AharonianAtoyan1983, kafexhiu2018b}. 

In Fig.~\ref{fig:ADAF-specSOLAR1}, we demonstrate the impact that this effect has on the spectrum of $\pi^0$-decay $\gamma$-rays in the case of the ADAF accretion regime. Aside from the "nominal" Maxwellian distribution, two other proton distributions have been considered: a Maxwellian distribution with a sharp cutoff at $4\,kT_{\rm i0}$ and a Maxwellian distribution with a suprathermal tail with a power-law index $p=2.0$ for energies above $4\,kT_{\rm i}$. At energies above a few 100 MeV, the corresponding $\gamma$-ray luminosities are significantly suppressed or enhanced compared to the luminosity corresponding to the Maxwellian distribution.

\section{Evaporation of neutrons}

\begin{figure}
\centering
\includegraphics[scale=0.45]{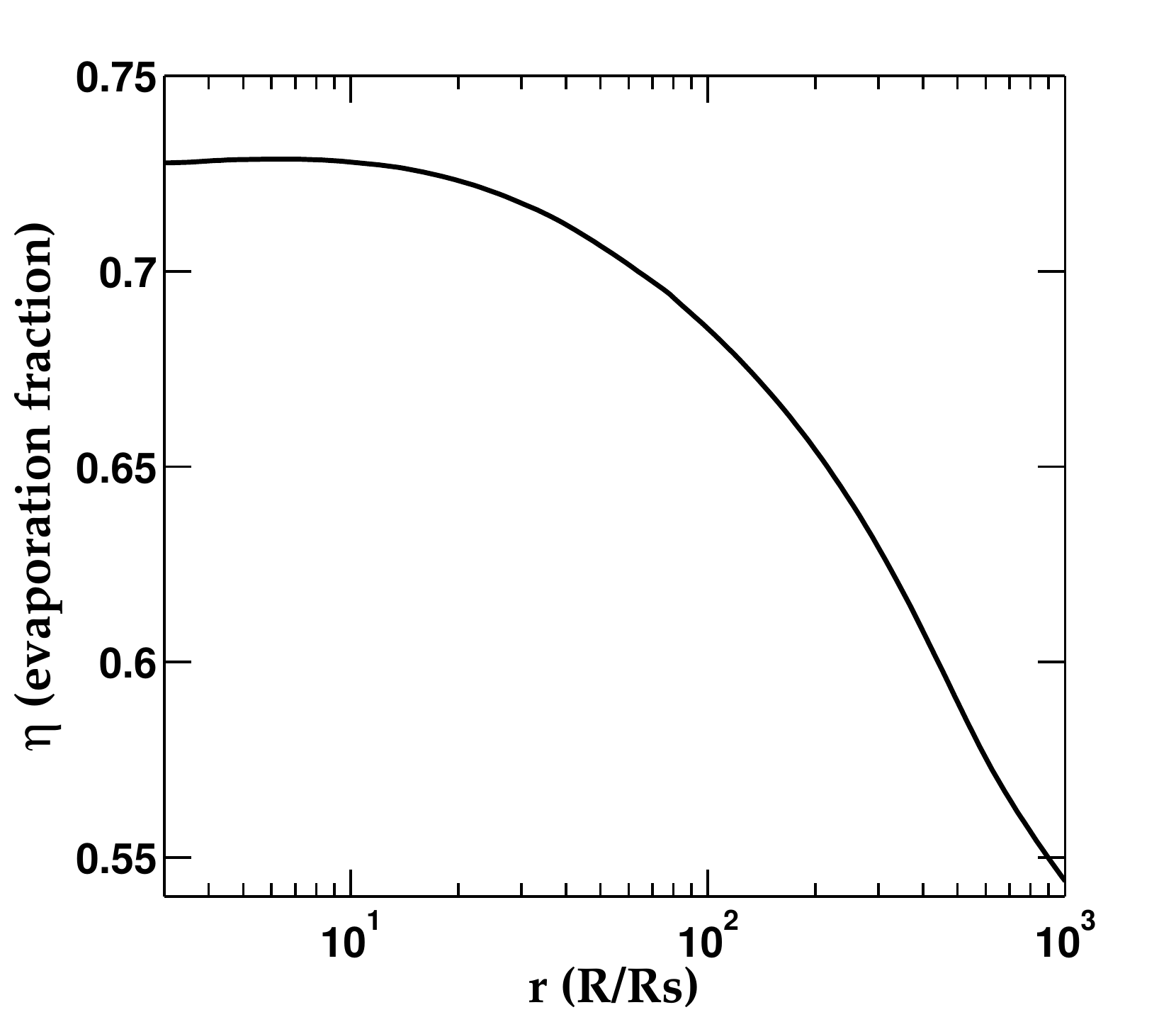}
\caption{Neutron evaporation efficiency $\eta$ as a function of the radius calculated for the ADAF model. This function is universal; it  does not depend strongly on the details of accretion.} 
\label{fig:evapor}
\end{figure}

Neutrons are not bounded by the electric and magnetic fields of the accretion plasma. If their energy -- the sum of thermal and rotation energies of the accretion disk -- is larger than the gravitational binding energy, neutrons can ``evaporate'' from the disk and reach the companion star before decaying \citep{Aharonian1984, Jean2001, Guessoum2002}. The ADAF predicts very hot plasma; therefore neutrons can easily escape from the disk almost isotropically. We calculated the neutron evaporation efficiency $\eta$ in the framework of the ADAF model by adopting the calculations described in \cite{Aharonian1984}, also considering the reduction of neutrons along the accretion disk. Figure~\ref{fig:evapor} shows $\eta$ as a function of the accretion radius. We find that the $\eta=f(r)$ profile is universal and does not depend on the accretion parameters. This follows from the fact that the ADAF ion temperature is equal to the virial temperature independent of the accretion parameters.

The primary source of neutrons in the accretion plasma is the destruction of the CNO and heavier nuclei. In addition, ADAF can produce neutrons through the $p+p\to p+n+\pi^+$ channel converting protons into neutrons. The neutrons are primarily produced in the $r\lesssim10$ region of ADAF.  Although this channel is less significant than that of nuclear destruction processes, the $p+p\to p+n+\pi^+$ channel ensures that ADAF can produce neutrons efficiently even if its accretion plasma is composed of merely hydrogen. Similar to the $\pi^0\to2\gamma$ emissivity, the neutron production rate through the process $p+p\to p+n+\pi^+$  can vary depending on the shape of the high energy tail of the plasma distribution function. 

\begin{figure}
\centering
\includegraphics[scale=0.45]{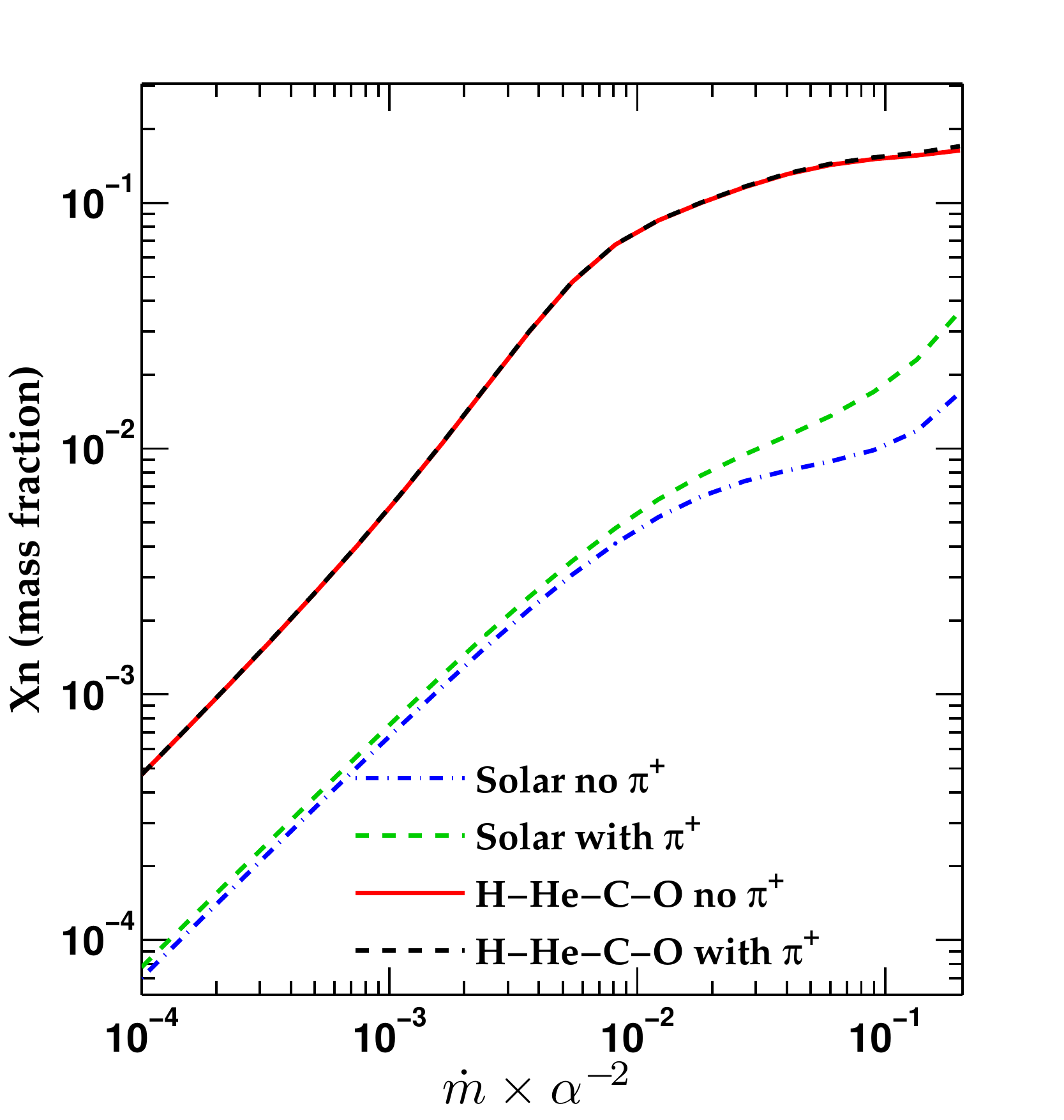}
\caption{Neutron mass fraction abundance at $r=3$ as a function of $\dot{m}\,\alpha^{-2}$. The magenta dash-dotted line and the cyan dashed line are calculated for the initial solar composition, obtained without and with the $p+p\to p+n+\pi^+$ contribution, respectively. The red line and the black dashed line are calculated for higher metallicity plasma with initial mass fraction for hydrogen, helium, carbon, and oxygen $X_p=X_\alpha=0.3$ and $X_{C12}=X_{O16}=0.2$. The black dashed line takes into account the contribution of the $p+p\to p+n+\pi^+$.}
\label{fig:Xn}
\end{figure}

The mass abundance scales of the neutrons such as $X_n \propto \rho/\varv \propto \dot{m}\,\alpha^{-2}$. The proportionality constant is a function of radius $r$. Figure~\ref{fig:Xn} shows the function $X_n=f(\dot{m}\,\alpha^{-2})$ for ADAF computed near the last stable orbit $r=3,$ which gives $X_n^{\rm max}$ throughout the disk. The neutron abundance is estimated for  two initial compositions: a solar composition and a heavier plasma with $X_p=X_\alpha=0.3$ and $X_{C}=X_{O}=0.2$. For comparison, $X_n$ is plotted with and without the  $p+p\to p+n+\pi^+$ contribution.

\section{Production of light elements}

\begin{figure}
\centering
\includegraphics[scale=0.45]{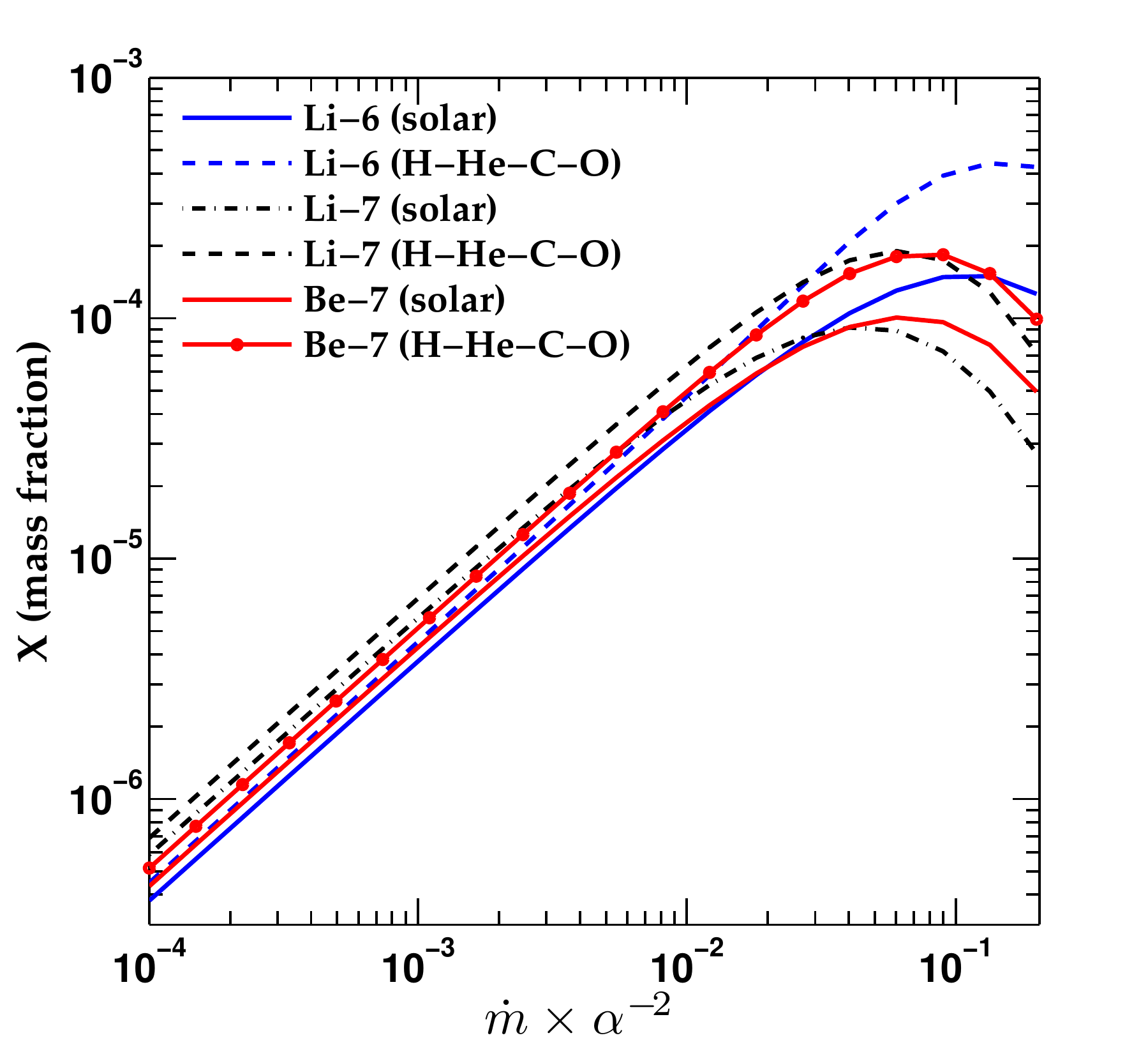}
\caption{Mass fraction abundances of $^6$Li (cyan), $^7$Li (black), and 
$^7$Be (red) at $r=3$ for ADAF as a function of $\dot{m}\,\alpha^{-2}$. The calculations are performed for two, solar, and "heavy" ($X_p=X_\alpha=0.3$ and $X_{C}=X_{O}=0.2$) initial plasma compositions.}\label{fig:ADAF-LiBe}
\end{figure}

Within the ADAF model, the accretion flow has a positive Bernoulli integral. Thus, this model can produce outflows \citep[e.g.,][]{Narayan1994a, Yi&Narayan1997, Blandford1999}. The outflows can carry some of the accretion material and inject it into the surrounding interstellar medium. The accretion material is richer with light elements such as Li and Be. Therefore, the expelled matter can enrich the interstellar medium with these elements in addition to the primordial (cosmological) source of Li and Be. Thus, the accretion disk outflows can, in principle, contribute to the solution of the so-called $^7$Li problem \citep[see, e.g.,][]{FieldsLi7prob2011}. Figure~\ref{fig:ADAF-LiBe} shows the mass fraction abundance of $^6$Li, $^7$Li, and $^7$Be as a function of $\dot{m}\;\alpha^{-2}$, for an ADAF plasma close to the last stable orbit $r=3$. The  calculations are performed for the solar and for a heavy ($X_p=X_\alpha=0.3$ and $X_{C}=X_{O}=0.2$) initial compositions.

\section{Discussion}

\subsection{Radial chemical evolution}

The radial chemical evolutions shown in Figs.~\ref{fig:ADAF-specSOLAR1}--\ref{fig:ADAF-specCO}, demonstrate that in the ADAF model the nuclear reactions play an important role starting from distances of $100~R_S$. The SLE plasma is colder, and, consequently, the impact of nuclear reactions is limited by the region $r\lesssim10$. The initial composition of the plasma is significantly modified as the nuclei move toward the inner parts of the accretion disk. In the ADAF model, all heavy nuclei are destroyed before they reach the last stable orbit $r=3$. As a consequence, the accretion matter that orbits near the black hole is composed of protons, neutrons, and light elements. In addition to ${\rm ^4He}$, the destruction timescale of which is longer than the accretion falling time, the plasma contains traces of ${\rm ^3He}$, D (at the level of $10^{-2}$), somewhat less T, as well as nuclei of the LiBeB group (at the level of $10^{-4})$

\subsection{$\gamma$-rays}

Gamma rays produced in nuclear interactions can be observed between $\approx 1$~MeV and 10~MeV. The region below 1~MeV is dominated by the thermal electron bremsstrahlung. The ADAF and SLE accretion models predict different gamma-ray spectra, basically because of differences in the temperature profiles. The plasma in the ADAF model is hotter. Therefore, $\gamma$-ray lines are expected to be broad because of larger Doppler broadening. Within 1 to 10~MeV, $\Delta E_\gamma/E_\gamma\approx 8$~\%.  The spectrum of the ADAF plasma continues above 10~MeV through the $\pi^0\to2\gamma$ emission. Its emissivity peaks around $E_\gamma\approx 100$~MeV and falls off exponentially at higher energies. The SLE plasma, on the other hand, produces narrower and more intensive nuclear lines and its $\gamma$-ray spectrum has a cutoff at $E_\gamma\approx10$~MeV.

The ADAF and SLE models have about the same efficiency of converting accretion energy into nuclear $\gamma$-ray lines, that is $L_N/L_{acc}< 5\times10^{-4}$. This efficiency decreases if nuclei are destroyed before reaching the inner parts of the accretion disk, where the temperature and density reach their maximum. Therefore, by increasing the plasma accretion rate or density, the nuclear destruction rates become large and, correspondingly,  the $\gamma$-ray efficiency drops. The maximum possible efficiency depends on the initial plasma composition. It is larger if the plasma is composed of heavier nuclei. In the ADAF model, the maximum luminosity for a solar initial composition is $\approx 10^{33}\times (m/10)~\rm{erg~s^{-1}}$.  For a heavier composition, for example, with the mass abundances  $X_p=X_\alpha=0.3$ and $X_{C12}=X_{O16}=0.2$, the luminosity is larger,  $L_N^{max} \approx 10^{34}\times (m/10)~\rm{erg~s^{-1}}$. We note that this estimate exceeds, by an order of magnitude, the result published by  \cite{Yi&Narayan1997}. 

The thermal $\pi^0$-meson production efficiency in the  ADAF model is on the order of $L_\pi/L_{acc}< 10^{-2}$. Unlike the nuclear $\gamma$-ray lines, the $\pi^0$-meson production efficiency has a strong dependence on the shape of the high energy tail of the plasma distribution function.   The simultaneous observations of nuclear $\gamma$-ray lines and $\pi^0$-decay $\gamma$-rays can offer a unique opportunity to reconstruct the plasma energy distribution function, which can be used to discriminate between different accretion regimes. While, the nuclear $\gamma$-ray lines contain information about the central part of the plasma distribution function, the $\pi^0\to2\gamma$ spectra provide information about its high energy tail.

The detectability of nuclear radiation from hot accretion flows depends on the accretion model. The most favorable conditions can be realized for the ADAF regime when the expected energy fluxes in the 1--10 MeV and 50--200 MeV bands from a source located at the distance $d$ can be as high as $10^{-11} (m/10) (d/{\rm 1kpc})^{-2} \,{\rm erg/cm^2\,s}$ and $3 \times 10^{-9} (m/10) (d/{\rm 1kpc})^{-2} \,{\rm erg/cm^2\,s}$, respectively (see section \ref{sec:Luminos}). The flux sensitivities of the planned $\gamma$-ray mission e-ASTROGAM \citep{astrogam} from the most promising galactic black hole binaries, such as Cyg X-1 ($m\approx15$ and  $d\approx2$~kpc) are significantly below the above fluxes, especially around 100~MeV. Therefore, e-ASTROGAM can provide meaningful probes of the ion temperature and accretion regimes through nuclear $\gamma$-rays from these objects.

\subsection{Evaporation of neutrons} 

Figure~\ref{fig:evapor} shows the neutron evaporation efficiency $\eta$ in the ADAF model as a function of the accretion disk radius. The $\eta=f(r)$ profile is universal for $r<100$ because it depends mainly on the plasma temperature, which is equal to the virial temperature for a wide range of the accretion disk parameters. We find that about 70~\% of the neutrons can escape from the accretion disk. 

The main source of neutrons are the CNO and heavier nuclei destruction. In addition, ADAF can effectively produce neutrons through the $p+p\to p+n+\pi^+$ channel that converts protons into neutrons. Like the $p+p\to \pi^0\to2\gamma$ emission, neutrons also originate from the central parts of the accretion disk ($r\lesssim10$) and their production rate depends on the shape of the high energy tail of the plasma distribution function.  Figure~\ref{fig:Xn} shows the maximum neutron fraction in the ADAF model that was  obtained at $r=3$ as a function of accretion parameters. 

Using the evaporation efficiency and the neutron content at different distances, we find that the ADAF can expel up to 15 percent of the accretion mass flow in the form of neutrons. Some of these neutrons can reach the companion star before decaying. These neutrons are cooled down and then captured by protons  in the atmosphere and produce the 2.22~MeV $\gamma$-rays that can serve as a signature of the discussed scenario \citep[see, e.g.,][]{Aharonian1984, Jean2001}.

For consistency, we included the neutron evaporation process in our nuclear reaction network solution. We found no significant effect on the nuclear abundances and $\gamma$-ray luminosity of the disk.

\subsection{Production of light elements}

Light elements are products of spallation of the CNO and heavier nuclei, as well as the synthesis of, for example, $p+\alpha$ and $\alpha+\alpha$ reactions. Figure~\ref{fig:ADAF-LiBe} shows the mass fraction abundances of $^6$Li, $^7$Li, and $^7$Be as a function of $\dot{m}/\alpha^2$ near the last stable orbit for the ADAF accretion disk. For a fixed radius, these abundances are proportional to $\dot{m}/\alpha^2$. However, for the accretion rates close to the critical rate, the destruction of $^6$Li, $^7$Li, and $^7$Be becomes important and therefore their abundances are reduced; see Fig.~\ref{fig:ADAF-LiBe}. \cite{Yi&Narayan1997} estimated lithium production in ADAF; their results are about an order of magnitude larger than the results we obtained via the nuclear reaction network. The reason for this discrepancy could be related to the neglection of Li destruction in their calculations.  

Because of the positive Bernoulli integral, ADAF can expel matter into the surrounding interstellar medium providing an additional source of light elements. Detailed quantitative studies of this process could contribute to the solution of the so-called problem of primordial $^7$Li \citep[see, e.g.,][]{FieldsLi7prob2011}.

\section{Summary and conclusions}

Using a large nuclear reaction network and the recent studies of the $\pi$-meson production cross sections at low energies, we computed the radial evolution of the chemical composition of the accreting matter and the $\gamma$-ray luminosities expected within the ADAF and SLE accretion disk models. Nuclear reactions are important within $100R_s$. The content of CNO and heavier nuclei decreases as the plasma moves toward the black hole event horizon. In the ADAF scenario, the plasma is sufficiently hot to destroy heavy nuclei before reaching the last stable orbit at $r=3$. The efficiency of converting accretion energy into nuclear $\gamma$-ray lines is $L_N/L_{acc}< 5\times10^{-5}$ for a solar composition and can be increased up to $10^{-3}$ for heavier compositions. The ADAF model can produce $\pi^0$-decay $\gamma$-rays with an efficiency as high as $L_\pi/L_{acc}< 10^{-2}$. This efficiency is sensitive to the high energy tail of the plasma distribution function. The ADAF accretion disk can evaporate about 70~\% of its neutron content with a rate that can be as large as 15~\% of the accretion rate. The ADAF model can also produce a considerable amount of elements from the LiBeB group that can be injected into the surrounding interstellar medium and alter its chemical composition.

\begin{acknowledgements}
The authors are grateful to the members of the High Energy Astrophysics Theory Group for the very fruitful discussions. The authors are appreciative for the grant from the Russian Science Foundation 16-12-1044, the NSF grant AST-1306672, DoE grant DE-SC0016369, and NASA grant 80NSSC17K0757.
\end{acknowledgements}

\bibliographystyle{aa}
\bibliography{refs}

\begin{thebibliography}{34}
\expandafter\ifx\csname natexlab\endcsname\relax\def\natexlab#1{#1}\fi

\bibitem[{{Abramowicz} {et~al.}(1995){Abramowicz}, {Chen}, {Kato}, {Lasota}, \&
  {Regev}}]{Abramowicz1995}
{Abramowicz}, M.~A., {Chen}, X., {Kato}, S., {Lasota}, J.-P., \& {Regev}, O.
  1995, \apjl, 438, L37

\bibitem[{{Abramowicz} {et~al.}(1988){Abramowicz}, {Czerny}, {Lasota}, \&
  {Szuszkiewicz}}]{Abramowicz1988}
{Abramowicz}, M.~A., {Czerny}, B., {Lasota}, J.~P., \& {Szuszkiewicz}, E. 1988,
  \apj, 332, 646

\bibitem[{{Abramowicz} {et~al.}(2000){Abramowicz}, {Lasota}, \&
  {Igumenshchev}}]{Abramowicz2000}
{Abramowicz}, M.~A., {Lasota}, J.-P., \& {Igumenshchev}, I.~V. 2000, \mnras,
  314, 775

\bibitem[{{Aharonian} \& {Atoyan}(1983)}]{AharonianAtoyan1983}
{Aharonian}, F.~A. \& {Atoyan}, A.~M. 1983, JETP, 58, 1079

\bibitem[{{Aharonian} \& {Sunyaev}(1984)}]{Aharonian1984}
{Aharonian}, F.~A. \& {Sunyaev}, R.~A. 1984, \mnras, 210, 257

\bibitem[{{Aharonian} \& {Syunyaev}(1987)}]{Aharonian1987}
{Aharonian}, F.~A. \& {Syunyaev}, R.~A. 1987, Astrophysics, 27, 413

\bibitem[{{Bisnovatyi-Kogan} {et~al.}(1980){Bisnovatyi-Kogan}, {Khlopov},
  {Chechetkin}, \& {Eramzhyan}}]{bisnovatyi1980}
{Bisnovatyi-Kogan}, G.~S., {Khlopov}, M.~Y., {Chechetkin}, V.~M., \&
  {Eramzhyan}, R.~A. 1980, \sovast, 24, 716

\bibitem[{{Blandford} \& {Begelman}(1999)}]{Blandford1999}
{Blandford}, R.~D. \& {Begelman}, M.~C. 1999, \mnras, 303, L1

\bibitem[{{Chen} {et~al.}(1995){Chen}, {Abramowicz}, {Lasota}, {Narayan}, \&
  {Yi}}]{Chen1995}
{Chen}, X., {Abramowicz}, M.~A., {Lasota}, J.-P., {Narayan}, R., \& {Yi}, I.
  1995, \apjl, 443, L61

\bibitem[{{Dahlbacka} {et~al.}(1974){Dahlbacka}, {Chapline}, \&
  {Weaver}}]{dahlbacka1974}
{Dahlbacka}, G.~H., {Chapline}, G.~F., \& {Weaver}, T.~A. 1974, \nat, 250, 36

\bibitem[{{de Angelis} {et~al.}(2018){de Angelis}, {Tatischeff}, {Grenier},
  {McEnery}, {Mallamaci}, {Tavani}, {Oberlack}, {Hanlon}, {Walter}, {Argan}, \&
  et~al.}]{astrogam}
{de Angelis}, A., {Tatischeff}, V., {Grenier}, I.~A., {et~al.} 2018, Journal of
  High Energy Astrophysics, 19, 1

\bibitem[{{Dermer}(1986)}]{dermer1986}
{Dermer}, C.~D. 1986, \aap, 157, 223

\bibitem[{{Fields}(2011)}]{FieldsLi7prob2011}
{Fields}, B.~D. 2011, Annual Review of Nuclear and Particle Science, 61, 47

\bibitem[{{Fowler} {et~al.}(1967){Fowler}, {Caughlan}, \&
  {Zimmerman}}]{Fowler1967}
{Fowler}, W.~A., {Caughlan}, G.~R., \& {Zimmerman}, B.~A. 1967, \araa, 5, 525

\bibitem[{{Giovannelli} {et~al.}(1982){Giovannelli}, {Karakula}, \&
  {Tkaczyk}}]{giovannelli1982}
{Giovannelli}, F., {Karakula}, S., \& {Tkaczyk}, W. 1982, \actaa, 32, 121

\bibitem[{{Guessoum} \& {Jean}(2002)}]{Guessoum2002}
{Guessoum}, N. \& {Jean}, P. 2002, \aap, 396, 157

\bibitem[{{Ichimaru}(1977)}]{Ichimaru1977}
{Ichimaru}, S. 1977, \apj, 214, 840

\bibitem[{{Jean} \& {Guessoum}(2001)}]{Jean2001}
{Jean}, P. \& {Guessoum}, N. 2001, \aap, 378, 509

\bibitem[{{Kafexhiu} {et~al.}(2018){Kafexhiu}, {Aharonian}, \&
  {Barkov}}]{kafexhiu2018b}
{Kafexhiu}, E., {Aharonian}, F., \& {Barkov}, M. 2018, ArXiv e-prints
  [\eprint[arXiv]{1807.06079}]

\bibitem[{Kafexhiu {et~al.}(2012)Kafexhiu, Aharonian, \& Vila}]{kafexhiu2012}
Kafexhiu, E., Aharonian, F., \& Vila, G. 2012, Int. J. Mod. Phys., D21, 1250009

\bibitem[{{Kolykhalov} \& {Syunyaev}(1979)}]{kolykhalov&sunyaev1979}
{Kolykhalov}, P.~I. \& {Syunyaev}, R.~A. 1979, \sovast, 23, 189

\bibitem[{{Koning} {et~al.}(2005){Koning}, {Hilaire}, \& {Duijvestijn}}]{talys}
{Koning}, A.~J., {Hilaire}, S., \& {Duijvestijn}, M.~C. 2005, in American
  Institute of Physics Conference Series, Vol. 769, International Conference on
  Nuclear Data for Science and Technology, ed. R.~C. {Haight}, M.~B.
  {Chadwick}, T.~{Kawano}, \& P.~{Talou}, 1154--1159

\bibitem[{{Kozlovsky} {et~al.}(2002){Kozlovsky}, {Murphy}, \&
  {Ramaty}}]{Kozlovsky2002}
{Kozlovsky}, B., {Murphy}, R.~J., \& {Ramaty}, R. 2002, \apjs, 141, 523

\bibitem[{{Mahadevan} {et~al.}(1997){Mahadevan}, {Narayan}, \&
  {Krolik}}]{madehavan1997}
{Mahadevan}, R., {Narayan}, R., \& {Krolik}, J. 1997, \apj, 486, 268

\bibitem[{{Murphy} {et~al.}(2009){Murphy}, {Kozlovsky}, {Kiener}, \&
  {Share}}]{Murphy2009}
{Murphy}, R.~J., {Kozlovsky}, B., {Kiener}, J., \& {Share}, G.~H. 2009, \apjs,
  183, 142

\bibitem[{{Narayan} \& {Popham}(1993)}]{narayan1993}
{Narayan}, R. \& {Popham}, R. 1993, \nat, 362, 820

\bibitem[{{Narayan} \& {Yi}(1995{\natexlab{a}})}]{Narayan1995a}
{Narayan}, R. \& {Yi}, I. 1995{\natexlab{a}}, \apj, 444, 231

\bibitem[{{Narayan} \& {Yi}(1995{\natexlab{b}})}]{Narayan1995b}
{Narayan}, R. \& {Yi}, I. 1995{\natexlab{b}}, \apj, 452, 710

\bibitem[{Narayan \& Yi(1994)}]{Narayan1994a}
Narayan, R. \& Yi, I.-s. 1994, Astrophys.J., 428, L13

\bibitem[{{Oka} \& {Manmoto}(2003)}]{oka2003}
{Oka}, K. \& {Manmoto}, T. 2003, \mnras, 340, 543

\bibitem[{{Paczynski} \& {Bisnovatyi-Kogan}(1981)}]{paczinski1981}
{Paczynski}, B. \& {Bisnovatyi-Kogan}, G. 1981, \actaa, 31, 283

\bibitem[{{Shakura} \& {Sunyaev}(1973)}]{ShakuraSunyaev1973}
{Shakura}, N.~I. \& {Sunyaev}, R.~A. 1973, \aap, 24, 337

\bibitem[{{Shapiro} {et~al.}(1976){Shapiro}, {Lightman}, \&
  {Eardley}}]{Shapiro1976}
{Shapiro}, S.~L., {Lightman}, A.~P., \& {Eardley}, D.~M. 1976, \apj, 204, 187

\bibitem[{{Yi} \& {Narayan}(1997)}]{Yi&Narayan1997}
{Yi}, I. \& {Narayan}, R. 1997, \apj, 486, 363

\end{thebibliography}

\end{document}